\documentclass[aps,pre,preprint]{revtex4}
\usepackage[T1]{fontenc}
\usepackage{amsmath}
\usepackage[latin1]{inputenc}
\usepackage{graphicx}
\renewcommand{\upsilon}{{\sf v}}

\usepackage{epsfig}

\usepackage{color}

\begin{document}
\title{Bulk fluid phase behaviour of colloidal platelet-sphere
  and platelet-polymer mixtures}

\author{Daniel de las Heras}
\email{delasheras.daniel@gmail.com}
\affiliation{Theoretische Physik II, Physikalisches Institut,
 Universit{\"a}t Bayreuth, D-95440 Bayreuth, Germany}

\author{Matthias Schmidt}
\affiliation{Theoretische Physik II, Physikalisches Institut,
 Universit{\"a}t Bayreuth, D-95440 Bayreuth, Germany}

\date{3 August 2012, revised version: 5 September 2012, to appear in
  Phil. Trans. A}

\begin{abstract}
  Using a geometry-based fundamental measure density functional
  theory, we calculate bulk fluid phase diagrams of colloidal mixtures
  of vanishingly thin hard circular platelets and hard spheres.  We
  find isotropic-nematic phase separation with strong broadening of
  the biphasic region upon increasing the pressure. In mixtures with
  large size ratio of platelet and sphere diameter, there is also
  demixing between two nematic phases with differing platelet
  concentrations. We formulate a fundamental-measure density
  functional for mixtures of colloidal platelets and freely
  overlapping spheres, which represent ideal polymers, and use it to
  obtain phase diagrams. We find that for low platelet-polymer size
  ratio in addition to isotropic-nematic and nematic-nematic phase
  coexistence, platelet-polymer mixtures also display
  isotropic-isotropic demixing.  In contrast, we do not find
  isotropic-isotropic demixing in hard core platelet-sphere mixtures
  for the size ratios considered.
\end{abstract}

\maketitle

\section{Introduction \label{introduction}}

Understanding the equilibrium properties of colloidal systems and the
relationship between the microscopic properties, such as particle
shapes and sizes, and the macroscopic properties of a dispersion is
essential for the task of creating new materials with desired
characteristics. Here anisotropic colloids are relevant due to the
possible formation of liquid crystalline phases. Among them
platelet-like colloids are ubiquitous in nature; gibbsite or certain
clays are specific examples.

Despite their simplicity, hard-core models are very suitable
candidates to investigate the phase behaviour of colloidal
systems. Since Onsager's pioneering work on the isotropic-nematic
($IN$) phase transition in a fluid of rods \cite{NYAS:NYAS627}, a
wealth of studies has been carried out in order to elucidate the
equilibrium properties of anisotropic hard particles, including the
fluid of circular platelets with vanishing thickness using
Onsager-like theories \cite{F29777300084, ForsythJr197837} and
fundamental measure theory (FMT) \cite{PhysRevE.73.011409,
  0953-8984-19-32-326103}; see
Ref. \cite{doi:10.1080/00268970802032301} for a recent review of the
literature for platelets and
Refs.~\cite{roth10review,tarazona08review,lutsko10review} for reviews
of FMT.

During the last few years, binary colloidal mixtures have received
considerable attention. The addition of a second component induces
profound changes in the phase diagram with new phenomena arising such
as demixing, reentrant phase boundaries, as well as the emergence of
phase transitions due to the depletion mechanism.  Studies of binary
mixtures that involve platelets were conducted for mixtures of thick
and thin platelets \cite{doi:10.1021/jp0105894, 0953-8984-16-19-013},
binary platelets with different diameters \cite{PhysRevE.81.041401},
and mixtures of platelets and rods \cite{varga:7207, C1SM06838E}.

In this paper we use density functional theory (DFT) \cite{evans79},
and in particular a geometry-based fundamental measure free energy
functional \cite{PhysRevE.73.011409}, in order to investigate the bulk
fluid properties in mixtures of infinitely thin platelets and hard
spheres. Such mixtures have been previously considered by Harnau and
Dietrich using DFT \cite{PhysRevE.71.011504} and by Oversteegen and
Lekkerkerker by means of a free-volume scaled-particle approach
\cite{oversteegen:2470}. The authors restricted their analyses to
isotropic states of platelets. Both studies found coexistence between
an isotropic fluid rich in platelets and a solid phase rich in
spheres, as well as the stability of a phase transition between two
isotropic phases with differing compositions for small values of the
platelet-sphere size ratio. Both studies neglect the excluded volume
interaction between the platelets, which can certainly be a good
approximation if the density of platelets is very low. We will compare
to these findings below.  We compare also to results from an extended
Onsager theory with Parsons-Lee rescaling
\cite{PhysRevA.19.1225,lee:4972} which includes two-body correlations
only. In contrast, the geometry-based density functional that we use
here includes explicitly non-local higher-body correlations.  The
non-local structure of the free energy functional
\cite{PhysRevE.73.011409} is specifically tailored for the present
shapes. For general convex bodies much recent effort
\cite{hansengoos09convex,korden2012virial,korden2012beyond} was
devoted to improve Rosenfeld's original FMT for non-spherical shapes
\cite{PhysRevLett.63.980,rosenfeld94convex,rosenfeld95convex}.

Here we also study mixtures of platelets and non-adsorbing ideal model
polymers \cite{PhysRevE.62.5225}. Polymers are modelled following the
ideas of Asakura and Oosawa \cite{oosawa:1255}, and Vrij \cite{vrij},
i.e., they are treated as fully interpenetrable spheres that cannot
overlap with the colloidal platelets.  This is to be regarded only as
a minimal model, in particular for the platelet-polymer interaction,
as both softness and penetrability of a platelet into a polymer coil
are neglected.  We develop a fundamental measure density functional
theory by linearizing the density functional for hard-core mixtures in
polymer density \cite{PhysRevLett.85.1934,0953-8984-14-40-323}.

Experimental mixtures of sterically stabilized gibbsite platelets and
silica spheres have been studied very recently
\cite{0953-8984-23-19-194109}. The authors found phase separation into
platelet-rich and sphere-rich phases, driven by the repulsive
interactions between particles of different species. Columnar ordering
has been observed in charged mixtures of gibbsite platelets and silica
spheres \cite{C1SM06535A,doi:10.1021/la101891e}. Furthermore the
crystallization of silica spheres in presence of gibbsite platelets
(silica-coated) was analyzed in Ref.~\cite{PhysRevE.71.041406}.

The paper is organized as follows. In Sec. \ref{theory} we define the
model and outline the DFT. The results are described in section
\ref{results}: mixtures of colloidal platelets and spheres are
analyzed in Sec. \ref{colloidal}, in Sec. \ref{polymers} we present
the results for mixtures of colloidal platelets and ideal polymers,
along with a comparison between FMT theory and an extended Onsager
theory which is given in \ref{scaled}. Suggestions for future research
lines and conclusions are presented in Sec. \ref{conclusions}.

\section{Model and theory}\label{theory}
\begin{figure}
\epsfig{file=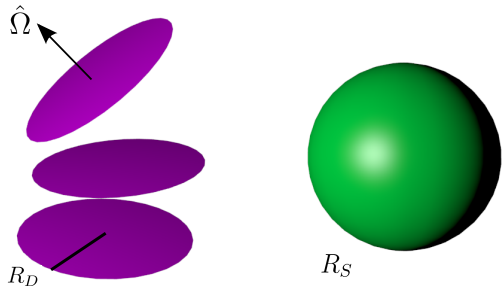,width=3.in,clip=}
\caption{(Online version in colour.) Schematic representation of the
  hard core platelet-sphere mixture. The model is a binary mixture of
  hard spheres of radius $R_S$ (right) and hard circular platelets of
  radius $R_D$ with vanishing thickness (left). The platelet
  orientation is described by a unit vector $\hat\Omega$ normal to the
  platelet surface.}
\label{fig1}
\end{figure}
We consider a binary mixture of $N_S$ hard spheres with radius $R_S$
and $N_D$ hard circular platelets with vanishing thickness and radius
$R_D$, see Fig.~\ref{fig1} for a schematic illustration. Here and
throughout species-dependent quantities are indicated by subscripts
$S$ (spheres) and $D$ (disks). The pair interaction potential between
any two particles is infinite if the particles overlap, and vanishes
otherwise.

\subsection{Density functional theory for colloidal platelet-sphere mixtures}\label{FMThardcores}
As is common, we split the Helmholtz free energy functional $F$ into
an ideal-gas part, $F_{id}$, and an excess contribution,
$F_{exc}$. Hence
\begin{equation}
F=F_{id}+F_{exc}.
\end{equation}
The ideal-gas free energy functional is exactly given by
\begin{equation}
\begin{aligned}
\beta F_{id}&=\int d^3r \int d\hat\Omega\rho_D(\vec r,\hat\Omega)[\ln(\rho_D(\vec r,\hat\Omega){\cal V}_D)-1]\\
&+\int d^3r \rho_S(\vec r)[\ln(\rho_S(\vec r){\cal V}_S)-1],
\end{aligned}
\end{equation}
where $\beta=1/k_BT$ is the inverse thermal energy, $k_B$ is the
Boltzmann constant, $T$ is absolute temperature, and ${\cal V}_i$ is
the (irrelevant) thermal volume of species $i$. The spatial integral
is over the system volume, $V$, and the angular integral is over the
unit sphere; here $\hat\Omega$ is a unit vector normal to the platelet
surface, cf.~Fig.~\ref{fig1}. The one-body distribution function of
platelets is denoted by $\rho_D(\vec r,\hat\Omega)$, and that of
spheres by $\rho_S(\vec r)$, the latter being trivially independent of
orientation due to the rotational symmetry of the spheres. In general
both distribution functions depend on position $\vec{r}$. In the
application to bulk fluid phase behaviour below we consider only
homogeneous states, which are independent of $\vec{r}$. Without loss
of generality we introduce an orientational distribution function for
the platelets, $\Psi_D(\vec r,\hat \Omega)$, via $\rho_D(\vec
r,\hat\Omega)=\rho_D(\vec r)\Psi_D(\vec r,\hat \Omega)$. Here
$\rho_D(\vec r)$ is the number density of platelets, which gives the
differential number of platelets with any orientation in the
differential volume element at $\vec{r}$.

The excess part of the free energy functional, $F_{exc}$, that
accounts for the excluded volume interactions between all particles,
is approximated by a geometry-based density functional, which has the
structure
\begin{equation}
\beta F_{exc}=\int d^3r\Phi(\{n_\nu^i\}),\label{excess}
\end{equation}
where the reduced free energy density, $\Phi$, is a function of a set
of weighted densities, $\{n_\nu^i\}$, where $i=S,D$ labels the
species, and $\nu$ the type of weight function. The free energy
density can be grouped into three different contributions,
$\Phi=\Phi_S+\Phi_D+\Phi_{SD}$. Here $\Phi_S$ describes sphere-sphere
interactions only. Both $\Phi_D$ and $\Phi_{SD}$ describe coupling of
spheres and platelets, with $\Phi_D$ containing the terms that
generate the FMT functional for pure platelets
\cite{PhysRevE.73.011409} in the limit of vanishing density of
spheres.  In detail, the reduced excess free energy for the
interaction between spheres, $\Phi_S$, is taken to be of the original
Rosenfeld form \cite{PhysRevLett.63.980},
\begin{equation}
\begin{aligned}
\Phi_S &= -n_0^S\ln(1-n_3^S)+\frac{n_1^Sn_2^S-\vec n_{\upsilon1}^S\cdot\vec n_{\upsilon2}^S}{1-n_3^S} \\
&+\frac{(n_2^S)^3-3n_2^S\vec n_{\upsilon2}^S\cdot\vec n_{\upsilon2}^S}{24\pi(1-n_3^S)^2}.
\end{aligned}
\end{equation} 
Here and in the following two equations we drop the spatial dependence
on $\vec{r}$ of the weighted densities for the sake of notational
convenience.  The interactions between platelets are described by
\begin{equation}
\begin{aligned}
\Phi_D &= \int d\hat\Omega
\frac{n_1^{DD}(\hat\Omega)n_2^D(\hat\Omega)}{1-n_3^S} \\
&+\int d\hat\Omega\int d\hat\Omega'\frac{n_2^{DDD}(\hat\Omega,\hat\Omega')n_2^D(\hat\Omega)n_2^D(\hat\Omega')}{24\pi(1-n_3^S)^2}.\label{phiD}
\end{aligned} 
\end{equation}
The first term in Eq. (\ref{phiD}) represents the interaction between
two platelets. It reduces to the exact second order contribution of
the virial series for the free energy of platelets if there are no
spheres in the mixture ($\rho_S=0$). The second term corresponds to
the interaction between three platelets (see
Ref. \cite{PhysRevE.73.011409,PhysRevE.81.041401} for details). The
factors proportionals to $(1-n_3^S)$ in the denominators account for
the volume occupied by the spheres in the system.

The further contribution to the reduced excess free energy due to the
interaction between platelets and spheres is
\begin{equation}
\begin{aligned}
\Phi_{SD} &= -\int d\hat\Omega n_0^D(\hat\Omega)\ln(1-n_3^S)\\
&+\int d\hat\Omega\frac{n_1^Sn_2^D(\hat\Omega)+n_2^{SD}(\hat\Omega)n_1^D(\hat\Omega)-\vec n_{\upsilon2}^{SD}(\hat\Omega)\cdot\vec n_{\upsilon1}^{D}(\hat\Omega)}{1-n_3^S}\\
&+\int d\hat\Omega\frac{[n_2^{SD}(\hat\Omega)n_2^{SD}(\hat\Omega)-\vec n_{\upsilon2}^{SD}(\hat\Omega)\cdot\vec n_{\upsilon2}^{SD}(\hat\Omega)]n_2^D(\hat\Omega)}{8\pi(1-n_3^S)^2}\\
&+\int d\hat\Omega\int d\hat\Omega'\frac{n_2^{SDD}(\hat\Omega,\hat\Omega')n_2^D(\hat\Omega)n_2^D(\hat\Omega')}{8\pi(1-n_3^S)^2}.
\end{aligned}
\end{equation}

All weighted densities are obtained by convolving the density with
specific weight functions. For hard spheres the scalar weighted
densities are:
\begin{equation}
n_{\nu}^S(\vec r)=w_\nu^S(\vec r)*\rho_S(\vec r),\quad \nu=0,1,2,3
\label{nHSscalar}
\end{equation}
and the vectorial weight densities are given by:
\begin{equation} 
  \vec n_{\nu}^S(\vec r)=\vec w_\nu^S(\vec r)*\rho_S(\vec r),\quad \nu=\upsilon1,\upsilon2,
\label{nHSvectorial}
\end{equation}
where $*$ denotes the three-dimensional convolution $h(\vec r)*g(\vec
r)=\int d^3xh(\vec x)g(\vec x-\vec r)$. Here $w_{\nu}^S$ and $\vec
w_{\nu}^S$ are the Rosenfeld weight functions for hard spheres
\cite{PhysRevLett.63.980}:
\begin{eqnarray}
\begin{aligned}
w_3^S(\vec r) &=\Theta(R_S-|\vec r|),\\
w_2^S(\vec r) &=\delta(R_S-|\vec r|),\\
\vec w_{\upsilon2}^S(\vec r) &=w_2^S(\vec r)\frac{\vec r}{|\vec r|},
\end{aligned}
\end{eqnarray}
where $\Theta(\cdot)$ is the Heaviside step function and $\delta(\cdot)$ is Dirac's delta distribution. Further weight functions for hard spheres are $w_1^S(\vec r)=w_2^S(\vec r)/(4\pi R_S)$, $w_0^S(\vec r)=w_2^S(\vec r)/(4\pi R_S^2)$, and $\vec w_{\upsilon1}^S(\vec r)=\vec w_{\upsilon2}^S(\vec r)/(4\pi R_S)$.

The sphere-platelet coupling is described by the weighted densities
\begin{eqnarray}
\begin{aligned}
n_{2}^{SD}(\vec r,\hat\Omega) &=w_2^{SD}(\vec r,\hat\Omega)*\rho_S(\vec r),\\
\vec n_{\upsilon2}^{SD}(\vec r,\hat\Omega) &=\vec w_{\upsilon2}^{SD}(\vec r,\hat\Omega)*\rho_S(\vec r),\\
n_{2}^{SDD}(\vec r,\hat\Omega,\hat\Omega') &=w_2^{SDD}(\vec r,\hat\Omega,\hat\Omega')*\rho_S(\vec r),
\end{aligned}
\end{eqnarray}
where the weight functions are
\begin{eqnarray}
\begin{aligned}
w_2^{SD}(\vec r,\hat\Omega) &=\frac4\pi\sqrt{1-(\vec r\cdot\hat\Omega/R_S)^2}w_2^S(\vec r),\\
\vec w_{\upsilon2}^{SD}(\vec r,\hat\Omega) &=4\frac{\vec r-(r\cdot\hat\Omega)\hat\Omega}{\pi R_S}w_2^S(\vec r),\\
w_2^{SDD}(\vec r,\hat\Omega,\hat\Omega') &=\frac8\pi\left|(\hat\Omega\times\hat\Omega')\cdot \vec w_{\upsilon2}^S(\vec r)\right|.
\end{aligned}
\end{eqnarray}

Finally, the weighted densities for platelets are
\begin{eqnarray}
\begin{aligned}
n_{\nu}^D(\vec r,\hat\Omega) &=w_\nu^D(\vec r,\hat\Omega)*\rho_D(\vec r,\hat\Omega),\quad \nu=0,1,2,\\
\vec{n}_{\upsilon1}^D(\vec r,\hat\Omega) &=\vec{w}_{\upsilon1}^D(\vec r,\hat\Omega)*\rho_D(\vec r,\hat\Omega),\\
n_1^{DD}(\vec r,\hat\Omega') &=\int d\hat\Omega w_1^{DD}(\vec r,\hat\Omega,\hat\Omega')*\rho_D(\vec r,\hat\Omega),\\
n_2^{DDD}(\vec r,\hat\Omega,\hat\Omega') &=\int d\hat\Omega''w_2^{DDD}(\vec r,\hat\Omega'',\hat\Omega,\hat\Omega')*\rho_D(\vec r,\hat\Omega''),
\end{aligned}
\end{eqnarray}
and their corresponding weight functions are
\begin{eqnarray}
\begin{aligned}
w_0^D(\vec r,\hat\Omega) &=\frac{1}{2\pi R_D}\delta(R_D-|\vec r|)\delta(\vec r\cdot\hat\Omega),\\
w_1^D(\vec r,\hat\Omega) &=\frac{1}{8}\delta(R_D-|\vec r|)\delta(\vec r\cdot\hat\Omega),\\
\vec w_{\upsilon1}^D(\vec r,\hat\Omega) &=\frac{\vec r}{R_D}w_1^D(\vec r,\hat\Omega),\\
w_2^D(\vec r,\hat\Omega) &=2          \Theta(R_D-|\vec r|)\delta(\vec r\cdot\hat\Omega),\\
w_1^{DD}(\vec r,\hat\Omega,\hat\Omega') &=\frac2{R_D}\left|\hat\Omega\cdot(\hat\Omega'\times\vec r)\right|w_1^D(\vec r,\hat\Omega),\\
w_2^{DDD}(\vec r,\hat\Omega,\hat\Omega',\hat\Omega'') &=\frac8{\pi}\left|\hat\Omega\cdot(\hat\Omega'\times\hat\Omega'')\right|w_2^D(\vec r,\hat\Omega).
\end{aligned}
\end{eqnarray}

The density functional described here is a special case of the
functional for ternary platelet-sphere-needle mixtures
\cite{PhysRevE.73.011409}. It includes contributions up to third-order
in platelet density. It reduces to the original Rosenfeld functional
\cite{PhysRevLett.63.980} in the case of a monodisperse fluid of hard
spheres, and gives results that are in good agreement with Monte Carlo
simulation data when describing the monodisperse fluid of vanishingly
thin platelets
\cite{PhysRevE.73.011409,PhysRevE.81.041401,cheung08isotropic}. Hence
we are confident that it can be applied successfully to the study of
binary mixtures of hard spheres and platelets.

\subsection{Density functional theory for mixtures of colloidal platelets and ideal polymers}\label{idealpolymers}
We use a model \cite{PhysRevE.62.5225} similar in spirit to the
colloid-polymer model of Asakura and Oosawa \cite{oosawa:1255}, and
Vrij \cite{vrij}, to study a mixture of colloidal platelets, which are
as above taken as infinitely thin circular disks, and ideal
polymers. The polymers are modeled by fully interpenetrable spheres of
radius $R_P$. These model polymers cannot overlap with the platelets.
We obtain an approximation for the free energy of this system by
linearizing the excess part of the free energy for the corresponding
hard core system. This strategy follows the derivation of the FMT for
the Asakura-Oosawa model from a corresponding binary hard sphere
mixture \cite{PhysRevLett.85.1934,0953-8984-14-40-323}. Hence we apply
the linearization to the hard core functional of Sec.~\ref{colloidal}
in order to obtain a free energy functional for the polymer-platelet
mixture, $F_{exc}^{PD}$, via
\begin{equation}
\begin{aligned}
F_{exc}^{PD}\left[\rho_D,\rho_P\right] &= F_{exc}\left[\rho_D,\rho_S=0\right]\\
&+\int d^3r \left.\frac{\delta F_{exc}}{\delta\rho_S(\vec{r})}\right|_{\rho_S=0}\rho_P(\vec{r}),
\end{aligned}
\end{equation} 
where $\rho_P(\vec{r})$ is the one-body polymer density distribution
and $F_{exc}$ is the free energy functional for the hard core mixture,
given in Eq. (\ref{excess}). $F_{exc}^{PD}$ can be expressed in terms
of a reduced free energy density
\begin{equation}
\beta F^{PD}_{exc}=\int d^3r\Phi_{PD}(\{n_\nu^i\}),
\end{equation}
where the reduced free energy, $\Phi_{PD}$, is a simple expression
linear in polymer density and with up to cubic contributions in
platelet density:
\begin{equation}
\begin{aligned}
\Phi_{PD} &= \int d\hat\Omega n_0^D(\hat\Omega)n_3^P(\hat\Omega)\\
&+\int d\hat\Omega\left(n_1^P(\hat\Omega)n_2^D(\hat\Omega)+n_2^{PD}(\hat\Omega)n_1^D(\hat\Omega)-\vec n_{\upsilon2}^{PD}(\hat\Omega)\cdot\vec n_{\upsilon1}^{D}(\hat\Omega)\right)\\
&+\int d\hat\Omega\int d\hat\Omega'\frac{n_2^{PDD}(\hat\Omega,\hat\Omega')n_2^D(\hat\Omega)n_2^D(\hat\Omega')}{8\pi}\\
&+\int d\hat\Omega
 n_1^{DD}(\hat\Omega)n_2^D(\hat\Omega)\left(1+n_3^P\right) \\
&+\int d\hat\Omega\int d\hat\Omega'\frac{n_2^{DDD}(\hat\Omega,\hat\Omega')n_2^D(\hat\Omega)n_2^D(\hat\Omega')}{24\pi}\left(1+2n_3^P\right).
\end{aligned} 
\end{equation}
We have suppressed the spatial dependence of the weighted densities in
the above expressions. The weighted densities that depend on the
polymer density (denoted by the letter $P$ in the superscript) are
those for hard spheres (\ref{nHSscalar}) and (\ref{nHSvectorial}), but
replacing the hard sphere colloid density, $\rho_S(\vec r)$, by the
polymer density $\rho_P(\vec r)$, and replacing the colloid radius,
$R_C$, by the polymer radius $R_P$.

\subsection{Spatially homogeneous fluids}\label{details}
In the following we restrict ourselves to considering spatially
homogeneous fluid states.  We use the sphere-platelet model, the
expressions for the polymer-platelet model being analogous.  The
density distributions are independent of the spatial coordinates and
\begin{eqnarray}
\begin{aligned}
\rho_S(\vec r) &=\rho_S = {\rm const},\\
\rho_D(\vec r,\hat\Omega) &=\rho_D\Psi_D(\hat\Omega), \text{ where } \rho_D={\rm const}
\end{aligned}
\end{eqnarray}
The total number of spheres, $N_S$, and that of platelets, $N_D$, are
obtained by integrating the number densities over the system volume
\begin{eqnarray}
\begin{aligned}
N_S &=\int d^3r \rho_S,\\
N_D &=\int d^3r\rho_D.\label{normaplatelets}
\end{aligned}
\end{eqnarray}
Eq. (\ref{normaplatelets}) implies that the orientational distribution
function of platelets is normalized according to
\begin{equation}
\int d\hat\Omega\Psi_D(\hat\Omega)=1,\label{norma}
\end{equation}
which is a different convention compared to that of
Ref.~\cite{PhysRevE.73.011409}.  As we do not expect biaxial
arrangements of the platelets to occur in bulk, we take the
orientational distribution function to depend only on the polar angle
$\theta$ with respect to the nematic director, hence
$\Psi(\hat\Omega)=\Psi(\theta)$. We use a parameter, $\Lambda$, and
prescribe the functional form of the orientational distribution
function as
\begin{equation}
\Psi_D(\theta)=\frac{\exp[\Lambda \mathrm{P_2}(\cos\theta)]}{\int d\hat\Omega \exp[\Lambda \mathrm{P_2}(\cos\theta)]}\label{odp},
\end{equation} 
satisfying by construction the normalization condition
(\ref{norma}). Here $\mathrm{P_2}(\cdot)$ is the second Legendre
polynomial, and $\Lambda$ determines the degree of orientational order
of the platelets. For isotropic states $\Psi_D(\theta)=1/4\pi$ is
obtained for $\Lambda=0$. We characterize the orientational order of
the platelets by $S_D$, the standard nematic order parameter:
\begin{equation}
S_D=\int d\hat\Omega \Psi_D(\theta)\mathrm{P_2}(\cos\theta).
\end{equation}

We have tested the quality of the parameterization (\ref{odp}) by
computing the $IN$ phase transition of the monodisperse fluid of
platelets, and comparing the coexisting densities and order parameter
to those obtained from free minimization
\cite{PhysRevE.73.011409}. The differences for the coexistence
densities between both methods are of the order of $1\%$, see Table
\ref{table} for a comparison.  The value of the nematic order
parameter at coexistence, $S_D^N$, deviates more strongly from the
result of free minimization. However, as its magnitude is unusually
small for the pure platelet system, this is a very sensitive quantity,
see the discussion in Ref.\ \cite{PhysRevE.81.041401}.  For
completeness, Table \ref{table} also gives the results from Onsager
theory and from Monte Carlo simulations.

\begin{table}[h]
\begin{center}
\vspace{.1in}
\begin{tabular}{|c|c|c|c|}
\hline
$\Psi_D(\theta)$ & $\rho_D^IR_D^3$ & $\rho_D^NR_D^3$ & $S_D^N$ \\
\hline
FMT, free minimization \cite{PhysRevE.81.041401}  & $0.419$ & $0.469$
& $0.533$ \\
FMT, parametrization (\ref{odp}) & $0.421$ & $0.462$ & $0.489$ \\
Monte Carlo  \cite{PhysRevE.57.4824} & $0.460$  & $0.498$ & $0.45-0.55$ \\
Onsager theory & $0.672$ & $0.858$ & $0.801$ \\
\hline
\end{tabular}
\caption{Coexistence values for the $IN$ phase transition of a
  monodisperse fluid of hard circular platelets with vanishing
  thickness. The values in the first row are from free minimization
  without an {\it a priori} form of the orientational distribution
  function \cite{PhysRevE.81.041401}. The second row shows the
  coexisting values using the parameterization (\ref{odp}). Monte
  Carlo data \cite{PhysRevE.57.4824} are presented in the third row.
  In the fourth row the results from Onsager theory are shown; these
  are identical to Parsons-Lee theory (Sec.~\ref{PL}), as the packing
  fraction vanishes in the pure system of platelets.  Here, $\rho_D^I$
  and $\rho_D^N$ are the coexistence densities at the $IN$ transition,
  and $S_D^N$ is the nematic order parameter in the coexisting nematic
  state.}\label{table}
\end{center}
\end{table}

In spatially homogeneous fluids, the weighted densities are obtained by integrating the weight functions over the spatial coordinates. For hard spheres:
\begin{eqnarray}
\begin{aligned}
n_0^S &=\rho_S,\\
n_1^S &=R_S\rho_S,\\
n_2^S &=4\pi R_S^2\rho_S,\\
n_3^S &=\frac43\pi R_S^3\rho_S,\\
\vec n_{\upsilon1}^S &= \vec n_{\upsilon2}^S=0.
\end{aligned}
\end{eqnarray}
The mixed weighted densities that couple spheres and platelets are:
\begin{eqnarray}
\begin{aligned}
n_2^{SD}(\hat\Omega) &=4\pi R_S^2\rho_S,\\
n_{\upsilon2}^{SD}(\hat\Omega) &=0,
\end{aligned}
\end{eqnarray}
\begin{widetext}
\begin{equation}
n_{2}^{SDD}(\theta,\theta')=\rho_S\int_{-R_S}^{R_S}dz\int_0^{2\pi}d\phi'\left\lbrace
  \begin{array}{l}
     32\pi|t| \text{ if } t^2 > s^2 \\
     64\left(\sqrt{s^2-t^2}+t\arcsin\frac ts \right) \text{ otherwise}, \\
  \end{array}
  \right.
  \label{eqn2sdd}
\end{equation}
where $t=z\sin\theta\sin\theta'\sin\phi'$, and
\begin{equation}
s=\sqrt{R_S^2-z^2}\sqrt{\sin^2\theta\cos^2\theta'+\cos^2\theta\sin^2\theta'-2\sin\theta\cos\theta\sin\theta'\cos\theta'\cos\phi'}.
\end{equation}
\end{widetext}
Due to our normalization (\ref{norma}), Eq.~(\ref{eqn2sdd}) above and
Eq.~(105) of Ref.~\cite{PhysRevE.73.011409} differ by a factor
$(4\pi)^2$.  The integrals over the azimuthal angles $\phi$ and
$\phi'$ have been performed in order to arrive at (\ref{eqn2sdd}), as
only the polar angle with respect to the nematic director is relevant
for uniaxial configurations. For platelets the bulk weight functions
reduce to
\begin{eqnarray}
\begin{aligned}
n_0^D(\hat\Omega) &=\rho_D\Psi_D(\hat\Omega),\\
n_1^D(\hat\Omega) &=\frac{\pi}4R_D\rho_D\Psi_D(\hat\Omega),\\
n_2^D(\hat\Omega) &=2\pi R_D^2\rho_D\Psi_D(\hat\Omega),\\
\vec n_{\upsilon1}^D(\hat\Omega) &=0.
\end{aligned}
\end{eqnarray}
Further contributions to the free energy due to the weighted densities
$n_1^{DD}$ and $n_2^{DDD}$ are the same as given in Eq. (13) and (27)
of Ref.~\cite{PhysRevE.81.041401} to which we refer the reader
directly. 

The angular integrals were calculated numerically using Gaussian
quadrature. We used $2000$ roots for the integrals over the azimuthal
angle and $25$ roots for the integrals in the polar angle. The
orientational distribution function does not depend on the azimuthal
angle, and hence the integrals over $\phi$ can be computed and stored
for subsequent use. We checked the accuracy of the numerical procedure
by computing selected coexisting points with $100$ roots for the
integrals over the polar angle. The differences were less than $1\%$.

\subsection{Parsons-Lee theory for colloidal platelet-sphere mixtures}\label{PL}
Parsons-Lee theory \cite{PhysRevA.19.1225,lee:4972} constitutes an
approximation where the second virial contribution to the excess free
energy is weighted with a prefactor in order to approximate the
effects of higher virial contributions. According to Parsons-Lee
theory the excess free energy for a spatially homogeneous binary
mixture of hard bodies can be written as (see
e.g.~\cite{cinacchi:234904}):
\begin{equation}
\frac{\beta F_{exc}}V=\phi(\eta)\sum_{ij}\rho_i\rho_j\int d\hat\Omega\int d\hat\Omega'\Psi_i(\hat\Omega)\Psi_j(\hat\Omega')v_{ex}^{ij}(\Omega,\Omega'), 
\label{PLfunctional}
\end{equation}
where $v_{ex}^{ij}(\hat\Omega,\hat\Omega')$ is the excluded volume
between two particles of species $i$ and $j$ with orientations given
by $\hat\Omega$ and $\hat\Omega'$, respectively. The prefactor
$\phi(\eta)$ depends on the chosen reference system. Here $\eta$ is
the total packing fraction of the mixture $\eta=\sum_i\rho_iv_i$,
where $v_i$ is the particle volume of species $i$. Our reference
system is a pure fluid of hard spheres, and for the present model
$\eta=v_S\rho_S$, as the platelet volume vanishes.  Using the
Carnahan-Starling equation of state, the prefactor in the excess free
energy is
\begin{equation}
  \phi(\eta)=\frac{4-3\eta}{8(1-\eta)^2}.
  \label{CS}
\end{equation} 
The orientational distribution function for spheres is
$\Psi_S(\hat\Omega)=1/4\pi$, and for platelets is given in
Eq. (\ref{odp}). The excluded volume between two spheres is
\begin{equation}
v_{ex}^{SS}(\hat\Omega,\hat\Omega')=\frac{32}3\pi R_S^3,
\end{equation}
between platelets and spheres
\begin{equation}
v_{ex}^{DS}(\hat\Omega,\hat\Omega')=2\pi R_SR_D^2+\pi^2R_DR_S^2+\frac43\pi R_S^3,
\end{equation}
and, finally, between two platelets
\begin{equation}
v_{ex}^{DD}(\hat\Omega,\hat\Omega')=4\pi R_D^3\sin\gamma,
\end{equation}
where $\gamma$ is the angle between $\hat\Omega$ and $\hat\Omega'$.
For pure platelets $\eta=0$, and hence (\ref{CS}) reduces to
$\phi(\eta)=1/2$, which renders (\ref{PLfunctional}) identical to the
(Onsager) second-virial functional.

\subsection{Coexistence conditions}
In what follows we denote the composition of the mixture, $x$, by the
molar fraction of hard spheres, {\it i.e.}, $x=\rho_S/\rho$, where
$\rho=\rho_S+\rho_D$ is the total density. The molar fraction of
platelets is then simply $\rho_D/\rho=1-x$.  The equilibrium
properties of the mixture are obtained by minimizing the Gibbs free
energy per particle $g=F/N_T+p/\rho$ at constant composition $x$,
pressure $p$, and temperature $T$. Here $F$ is the total Helmholtz
free energy, and $N_T=N_S+N_D$ is the total number of particles. We
use a standard Newton-Raphson method to minimize $g$ as a function of
$\rho$ and $\Lambda$. Binodal lines are located by a common-tangent
construction on $g(x)$, which is equivalent to the equality of
chemical potentials of both species at the coexisting values of $x$
and $\rho$.  Thermal and mechanical equilibrium are satisfied in
advance by fixing both temperature and pressure.

Correspondingly, for the platelet-polymer mixture, $x=\rho_P/\rho$,
and $\rho=\rho_P+\rho_D$. The coexistence conditions are those given
above with colloidal spheres quantities exchanged by the respective
polymer quantity.

\section{Results}\label{results}
The pure fluid of hard platelets with vanishing thickness undergoes a
first order $IN$ phase transition as a function of density. Table
\ref{table} summarizes the values for the coexisting densities and for
the nematic order parameter according to the present FMT. The results
are in good agreement with Monte Carlo simulation data
\cite{marechal:094501,PhysRevLett.49.1089}. Columnar and solid phases
appear only in the limit of infinite density in a pure system of
platelets with vanishing thickness \cite{PhysRevE.57.4824}. In a
monodisperse fluid of hard spheres there is a well-known first-order
fluid-solid phase transition by increasing the density. We have not
studied spatially ordered phases, but give some comments about the
stability of spatially homogeneous phases with respect to hard sphere
crystallization in Sec. \ref{conclusions}.

\subsection{Mixtures of colloidal platelets and spheres}\label{colloidal}
\begin{figure*}
\epsfig{file=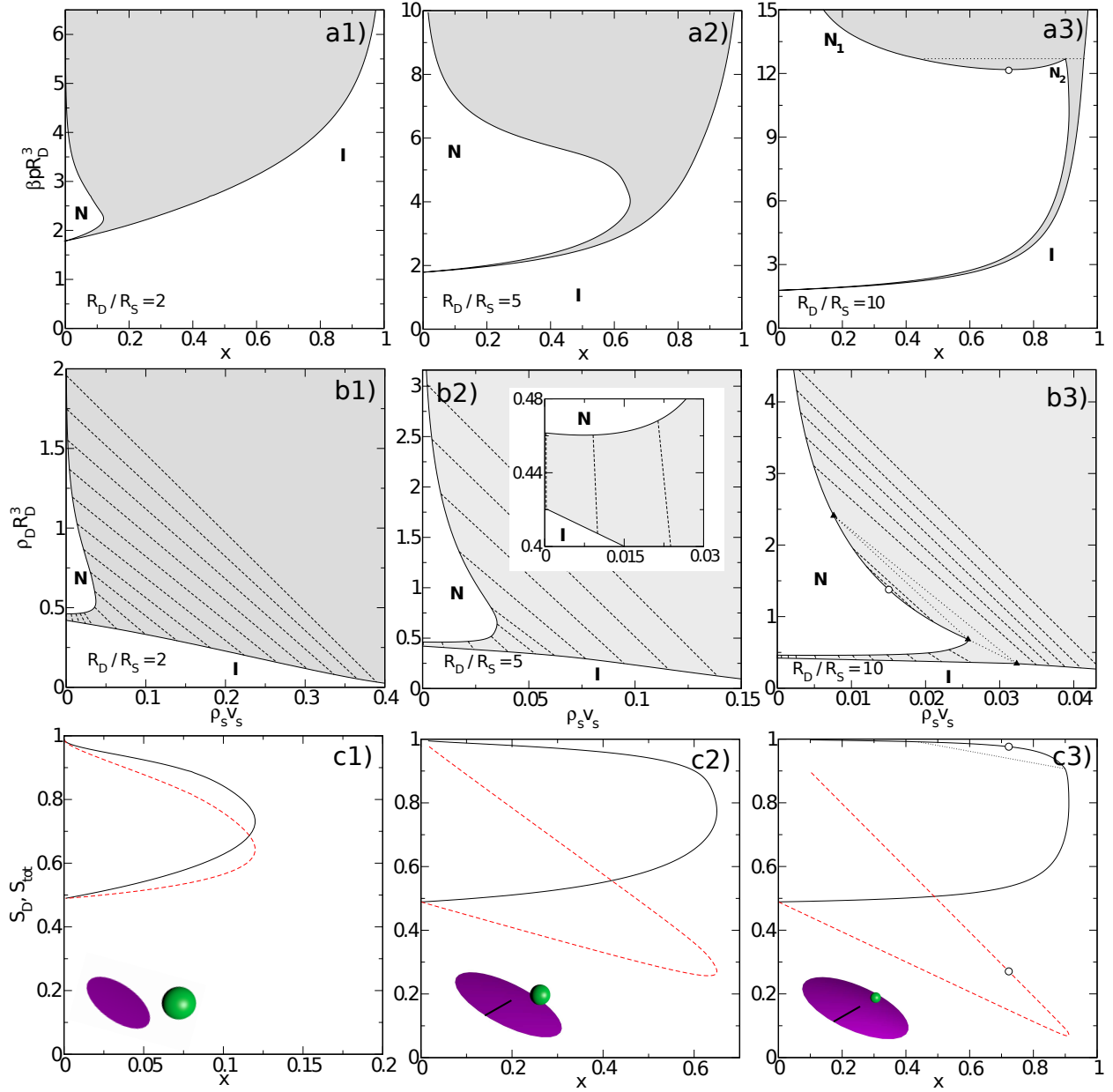,width=6.5in,clip=}
\caption{(Online version in colour.) Phase diagrams of binary mixtures
  of spheres and platelets with three different values of the
  platelet-sphere size ratio: $R_D/R_S=2$ (first column), $5$ (second
  column), and $10$ (third column). For each mixture we show the
  reduced pressure $\beta pR_D^3$ vs composition of spheres $x$ (first
  row), the reduced density of platelets $\rho_DR_D^3$ as a function
  of the packing fraction for spheres $\rho_S v_S$ (second row), and
  the nematic order parameters along the nematic side of the binodal
  (third row). The shadowed areas indicate two-phase regions. Empty
  circles represent critical points. Dotted lines represent triple
  points. The dashed lines in the second row are selected tie lines
  that connect two coexisting points on the binodal. The inset in
  panel (b2) is a zoom of the region close to the $IN$ phase
  transition of the pure fluid of platelets. The solid black (dashed
  red) lines in the third row represent the platelets' (total) nematic
  order parameter, $S_D$ ($S_T$). The insets in the third row are
  cartons showing the relative size between species.}
\label{fig2}
\end{figure*}

In Fig.~\ref{fig2} we present the phase behaviour of mixtures of
platelets and spheres for three different mixtures with size ratios
$R_D/R_S=2$ (first column), $5$ (second column), and $10$ (third
column). We comment on the behaviour for smaller size ratios at the
end of the section.  In Fig.~\ref{fig2} (a1) we show the phase diagram
in the plane of reduced pressure, $\beta pR_D^3$, and molar fraction
of spheres, $x$, for a mixture with size ratio $R_D/R_S=2$. The $IN$
phase transition in the pure fluid of platelets ($x=0$) moves to
higher pressures and the fractionation becomes stronger when spheres
are added to the system. Upon increasing the pressure, the nematic
branch of the binodal initially moves to higher compositions, but then
bends back on itself approaching $x=0$. The nematic phase is stable
only when the molar fraction for spheres is very low.

In Fig.~\ref{fig2} (a2) we plot the phase diagram for the same mixture
in the plane of reduced density of disks, $\rho_DR_D^3$, and packing
fraction of spheres, $\rho_Sv_S$, where $v_S$ is the sphere
volume. The $IN$ biphasic region strongly broadens when the packing
fraction of spheres in the mixture is increased.  The orientational
order parameter of platelets, $S_D$, along the nematic side of the
binodal is shown in panel (a3) of Fig.~\ref{fig2}.  We also display
the total orientational order parameter of the mixture i.e., the
nematic order parameter of platelets weighted by the composition of
platelets:
\begin{equation}
S_{tot}=\rho_D/\rho S_D=(1-x)S_D. 
\end{equation}
As the composition of spheres at the $IN$ phase transition is low,
both orientational order parameters are similar to each other,
increasing along the nematic side of the binodal.

The second column of Fig.~\ref{fig2} shows the results for mixtures
with $R_D/R_S=5$, i.e., for increased relative size of the
platelets. The topology of the phase diagrams in both representations
is the same than in the above mixture. The main difference to the case
of $R_D/R_S=2$ is the strongly increased range of compositions for
which the nematic phase is stable. For example, in Fig.~\ref{fig2}
(a2) one can observe that, for a range of pressures, it is possible to
find a stable nematic phase even if the composition of the mixture is
$x\gtrsim 0.6$. The mechanism that induces the $IN$ phase transition is
different at low and at high pressures. At low pressures the
transition is mainly due to the gain in configurational entropy of the
platelets in the nematic phase, as the excluded volume between two
platelets is minimal if they are parallel to each other. In this
regime there is low partitioning between the isotropic and nematic
phases. At high pressures, however, the transition is driven by the
(unfavourable) excluded volume interactions between platelets and
spheres. As a result there is strong demixing between a nematic phase
rich in platelets and an isotropic phase mostly composed of spheres.

We focus next on the behaviour of $\rho_D$ along the nematic branch of
the binodal. For low size ratios it monotonically increases as spheres
are added to the mixture; but for $R_D/R_S\gtrsim2$, $\rho_D$ slightly
decreases when the packing fraction of spheres is low (see the inset
in panel (b2) of Fig.~\ref{fig2}). This behaviour appears to be
counterintuitive. However a reduction of the density of platelets at
which the nematic phase is stable in the mixture has recently been
observed in sedimentation experiments on sterically stabilized
mixtures of gibbsite platelets and silica spheres
\cite{0953-8984-23-19-194109} of size ratio $\approx6$. However, the
authors found a much stronger effect than that obtained in our
theory. The quantitative difference could be due to the influence of
gravity on the bulk phase diagram \cite{sedimentationpaper}. The
polydispersity of the components, the finite thickness of the
platelets, and further interactions between the particles (not purely
entropic as we are considering here) in the experimental setup could
also be relevant to explain the quantitative differences.
 
The last column in Fig.~\ref{fig2} shows results for the phase
behaviour of a mixture with size ratio $R_D/R_S=10$.  A prominent
alteration of the topology of the phase diagram is that in addition to
the $IN$ phase transition, there is also demixing between two nematic
phases with differing compositions. The nematic-nematic phase
coexistence region is bounded by a lower critical point and merges
with the $IN$ phase transition at an isotropic-nematic-nematic triple
point. At pressures above the triple point pressure there is strong
demixing between an isotropic phase rich in spheres and a nematic
phase mostly composed of platelets. The nematic-nematic demixing can
be viewed as being driven by a depletion attraction induced between
the platelets by the small spheres.

The behaviour of the orientational order parameters along the binodal
is shown in the third row of Fig.~\ref{fig2}. For small size ratios,
panel (c1), both the order parameter of platelets, $S_D$, and the
total order parameter, $S_{tot}$, increase along the nematic branch of
the binodal. However, in mixtures where the size ratio is high (panels
(c2) and (c3) of Fig.~\ref{fig2}) $S_D$ monotonically increases along
the binodal, but the total order parameter, $S_{tot}$, first decreases
to a minimum and then increases. This qualitative difference may be
observable experimentally.

The phase diagram of mixtures with size ratios $R_D/R_S<2$ (not shown)
are qualitatively similar to the results for $R_D/R_S=2$. In
particular, the topology remains unchanged and only $IN$ phase
separation occurs. This is in contrast to previous studies of mixtures
of infinitely thin platelets and hard spheres
\cite{oversteegen:2470,PhysRevE.71.011504}, which reported
isotropic-isotropic ($II$) demixing if the size ratio of the mixture
is small. In our DFT treatment, we do not find $II$ coexistence in the
range of size ratios $0.2 \leq R_D/R_S \leq 10$.

\subsection{Mixtures of colloidal platelets and ideal polymers}\label{polymers}
\begin{figure*}
\epsfig{file=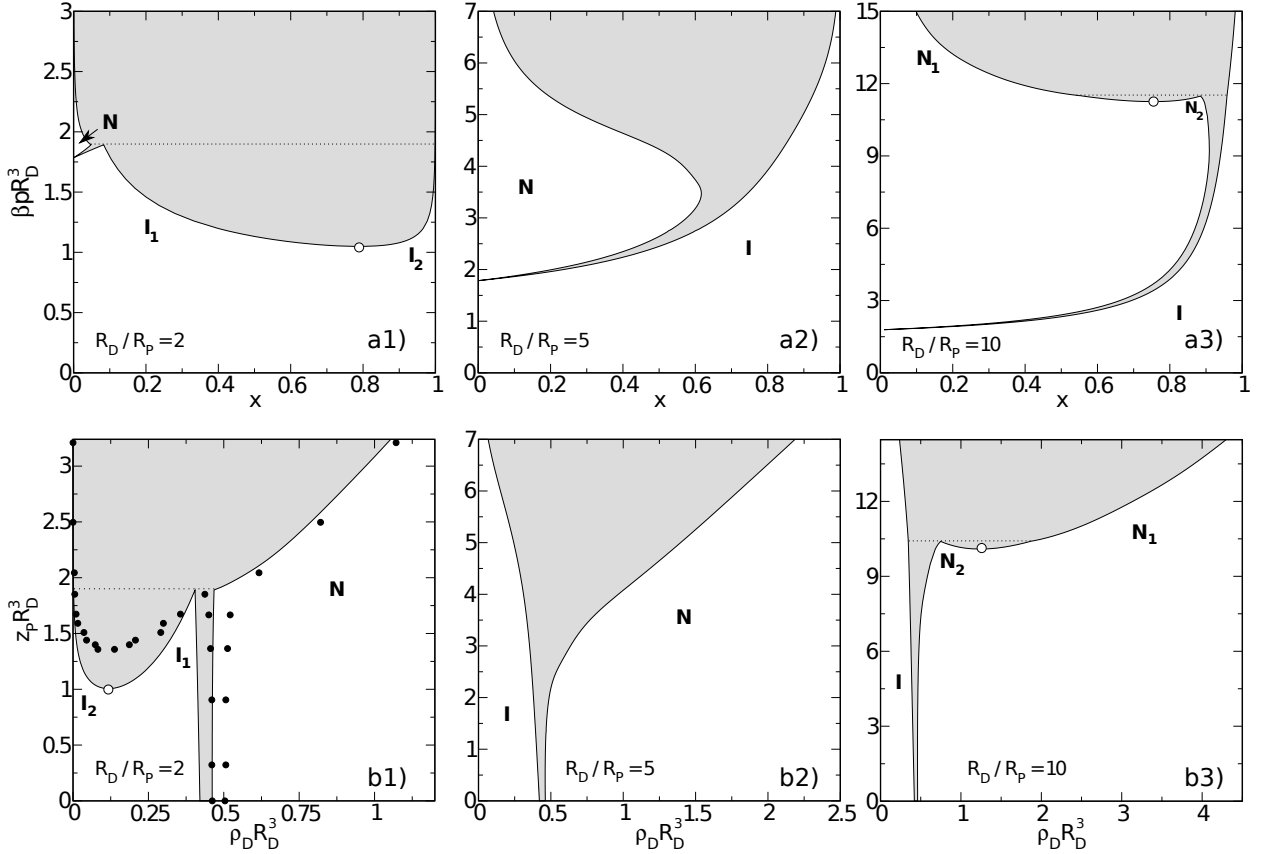,width=6.5in,clip=}
\caption{Phase diagrams of mixtures of colloidal platelets and ideal
  model polymers for size ratios $R_D/R_P=2$ (left), $R_D/R_P=5$
  (middle), and $R_D/R_P=10$ (right). Upper row: pressure, $\beta
  pR_D^3$, as a function of the composition of polymers, $x$. Bottom
  row: fugacity of the polymers, $z_PR_D^3$, vs density of platelets,
  $\rho_DR_D^3$. Empty circles present critical points. Dotted lines
  represent triple points. Full circles in panel (b1) indicate Monte
  Carlo simulation data \cite{PhysRevE.62.5225}.}
\label{figpolymers}
\end{figure*}
We turn to the analysis of mixtures of colloidal platelets and ideal
model polymers. Bates and Frenkel studied such mixtures using Monte
Carlo simulations and perturbation theory
\cite{PhysRevE.62.5225}. Following the ideas of Asakura, Oosawa and
Vrij \cite{oosawa:1255,vrij}, the polymers are modelled as freely
interpenetrable spheres that cannot overlap with the platelets due to
hard repulsion between the unlike species. A DFT for such mixture is
obtained by linearizing the excess free energy, Eq. (\ref{excess}),
around $\rho_S=0$ (see Sec. \ref{idealpolymers} for details).

The bulk fluid phase behaviour of mixtures of colloidal platelets and
non-adsorbing polymers is depicted in Fig.~\ref{figpolymers} for three
different values of the size ratio $R_D/R_P$.  The upper row shows the
phase diagram in the pressure-composition plane. The polymer fugacity,
$z_P=\exp(\beta\mu_P)/{\cal V}_P$, with $\mu_P$ the chemical potential
of the polymers, as a function of the density of platelets is
represented in the lower row.  Here the chemical potential of species
$i$ is obtained from $\partial F/\partial N_i=\mu_i$. Note that the
fugacity is independent of the (hence irrelevant) thermal volume
${\cal V}_P$ because the chemical potential, $\mu_p$ is shifted
according to the particular choice of ${\cal V}_P$.

The main difference to platelet-colloid mixtures is that mixtures of
platelets and ideal polymers display prominent demixing between two
isotropic states when the size ratio is low enough. In panel (a1) of
Fig.~\ref{figpolymers} we show the results for $R_D/R_P=2$. There is
strong demixing between two isotropic phases bounded by a lower
critical point. The $II$ phase separation occurs almost entirely at
pressures below the $IN$ pressure of the pure fluid of platelets. At
higher pressures the $II$ binodal merges with the $IN$ binodal via a
nematic-isotropic-isotropic triple point. This phase diagram differs
significantly from that of the hard core mixture with the same size
ratio shown in panel (a1) of Fig.~\ref{fig2}. The difference between
both systems lies in the sphere-sphere excluded volume
interactions. In a mixture of platelets and ideal polymers the gain in
accessible volume in the demixed $II$ state over-compensates the loss
of mixing entropy. This is not the case in mixtures of colloidal
platelets and spheres for the size ratios considered. It is
interesting to note that $II$ demixing has been found by computer
simulation in mixtures of platelets with finite thickness and
non-adsorbing polymers \cite{zhang:9947}.

In panel (b1) of Fig.~\ref{figpolymers} we show a comparison with
Monte Carlo simulation data (full circles) \cite{PhysRevE.62.5225} for
the same mixture ($R_D/R_P=2$). The agreement is reasonably good,
although DFT underestimates the $IN$ coexisting densities at low
polymer concentration and also the fugacity of the polymers along both
branches of the $II$ binodal and hence the location of the critical
point. This trend is similar to what was found for the AOV model
\cite{vink04jcp}.

The middle (right) column of Fig.~\ref{figpolymers} shows the phase
diagrams of a mixture with size ratio $R_D/R_P=5$ ($R_D/R_P=10$). The
behaviour is very similar to the corresponding mixtures of colloidal
spheres and platelets, depicted in panels (a2) and (a3) of
Fig.~\ref{fig2}. Such similarity might be expected because by
increasing the relative size of the platelets (in hard-core
platelet-sphere mixtures) the platelet-platelet and platelet-sphere
interactions dominate over the sphere-sphere interactions. Recall that
we do not consider the freezing transition of hard spheres, which
would dramatically change the phase behaviour between ideal polymers
and colloidal spheres at densities above the freezing transition.

Our results are in qualitative agreement with the experimental
observations. For example, $II$ coexistence has been found in
suspensions of sterically stabilized colloidal gibbsite platelets and
non-adsorbing polymers (with $R_D/R_P\approx2.9$) by Lekkerkerker {\it
  et al.} \cite{PhysRevE.62.5397}. The same study also showed a strong
broadening of the $IN$ two-phase region by increasing the polymer
concentration. Nematic-nematic demixing has been recently found
\cite{doi:10.1021/la8015595} in a related system of mixtures of
non-adsorbing polymers and positively charged platelets.

\subsection{FMT vs Parsons-Lee theory}\label{scaled}
\begin{figure}
\epsfig{file=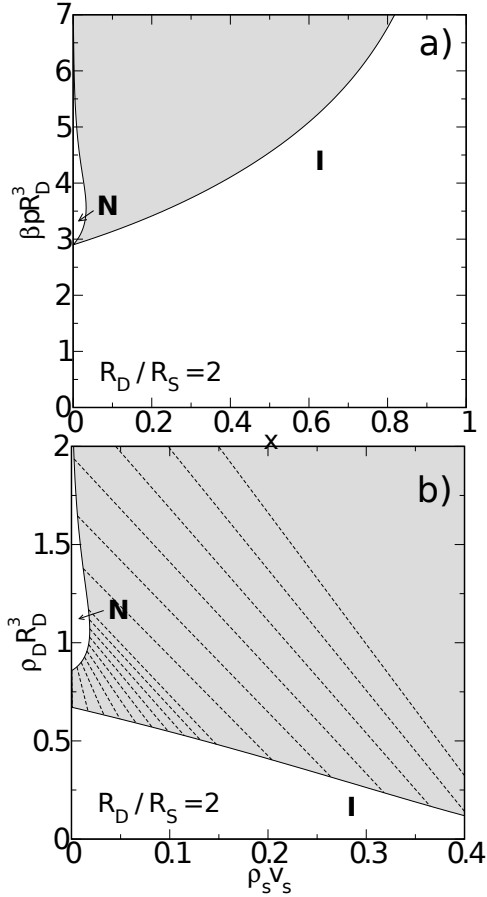,width=2.5in,clip=}
\caption{Bulk phase diagrams of a colloidal mixture of platelets and
  spheres according to Parsons-Lee theory. (a) Scaled pressure $\beta
  pR_D^3$ vs composition of spheres $x$. (b) Density of platelets
  $\rho_DR_D^3$ as a function of the packing fraction of spheres
  $\rho_Sv_S$. The size ratio is $R_D/R_S=2$.}
\label{figPL}
\end{figure}
The geometry-based DFT of Sec. \ref{FMThardcores} contains
contributions to the excess free energy that are of third order in
density.  In this section we compare the results from this theory to
those from an extended Onsager theory with Parsons-Lee rescaling that
explicitly includes only two-body interactions. For details about
Parsons-Lee theory, see Sec. \ref{PL}.  The phase diagram of a hard
core mixture with size ratio $R_D/R_S=2$ according to Parsons-Lee
theory is depicted in Fig.~\ref{figPL}. It is to be compared with
panels (a1) and (b1) of Fig.~\ref{fig2}, where corresponding results
obtain from FMT are shown. An immediately noticeable difference is in
the location of the $IN$ phase transition in the pure fluid of
platelets. In this limit ($x=0$) Parsons-Lee theory gives an unchanged
result as compared to Onsager theory due to the vanishing packing
fraction in the pure fluid of platelets. Hence Parsons-Lee theory
clearly overestimates the coexisting densities ($\rho_D^IR_D^3=0.67$
and $\rho_D^NR_D^3=0.86$). It also predicts a much too high nematic
order parameter at the $IN$ coexistence, $S_N=0.80$ (not shown),
compared to FMT and MC simulation.

For $R_D/R_S=2$ the topology of the phase diagram predicted by both
theories is the same. According to Parsons-Lee theory the partitioning
between isotropic and nematic phases is higher than in FMT. A similar
trend was observed comparing the current FMT and Onsager theory in
binary mixtures of hard platelets with different radius
\cite{PhysRevE.81.041401}.  Parsons-Lee theory predicts demixing
between two nematic phases in mixtures with high size ratio, as does
FMT. However, the minimal size ratio for phase segregation to occur
between two nematics (using Parsons-Lee theory) is $R_D/R_S\gtrsim15$,
considerably higher than the threshold for FMT, which is
$R_D/R_S\approx10$. One important difference is that Parsons-Lee
theory predicts $II$ demixing for low size ratios
$R_D/R_S\lesssim1$. Nevertheless, the $II$ phase boundary is located
at very high pressures and densities and is most likely metastable
with respect to segregation between an isotropic phase rich in
platelets and a solid phase rich in spheres.

\section{Conclusions}\label{conclusions}

We have investigated the fluid bulk phase behaviour of mixtures of
colloidal platelets and spheres using a geometry-based density
functional theory \cite{PhysRevE.73.011409}. The size ratio $R_D/R_S$
is the key parameter that controls the behaviour of the system. In
mixtures with $0.2\leq R_D/R_S\lesssim10$ we find only $IN$ phase
separation. We have identified two different mechanisms behind the
$IN$ phase transition. If the composition of spheres is low, the $IN$
transition is driven by the excluded volume interaction between the
platelets, similar to the mechanism in a pure fluid of platelets. In
this regime, the transition takes place between two phases with low
partitioning. As the composition of spheres is increased, the
interaction between dissimilar species becomes dominant, and the
(unfavourable) excluded volume interactions between spheres and
platelets drives strong demixing between an isotropic sphere-rich
phase and a nematic platelet-rich phase.

Mixtures with high size ratio, $R_D/R_S\gtrsim10$, also display
demixing between two nematics at different compositions. However
reliably describing such highly asymmetric mixtures is a very
difficult for any theoretical treatment, see the comparison of FMT and
simulation results for asymmetric binary hard sphere mixtures
\cite{herring06prl,PhysRevE.84.061136}.

The possible existence of demixing between two isotropic fluids in
mixtures of colloidal platelets and spheres remains an open
question. The FMT approach does not predict $II$ demixing. Previous
studies \cite{oversteegen:2470,PhysRevE.71.011504} reported $II$ phase
separation in mixtures were the size of both species is
similar. Nevertheless, in both studies the authors neglected excluded
volume interactions between platelets. This approximation is valid
only if the platelets density is very low.  In contrast our theory
includes platelet-platelet interactions, modelled via up to cubic
contributions in platelet density to the free energy. The bulk free
energy for isotropic states from the present DFT is the same as that
from scaled-particle theory \cite{Barker76}, which gives further
confidence into our results. Carrying out a careful simulation study
in order to shed further light on the existence of $II$ demixing in
the hard core mixture would be very valuable.

We have not considered spatially inhomogeneous phases in this
work. The pure fluid of hard spheres undergoes a fluid-solid phase
transitions. The freezing transition, according to Monte Carlo
simulations \cite{vega:154113,fortini:4-18-28-L02}, takes place at
packing fractions $\rho_Sv_S=0.49$ for the liquid and $0.54$ for the
solid (values which are higher than the packing fractions considered
in the present work). Hence, we expect our fluid phase diagrams to be
stable with respect to freezing of the hard spheres. For low size
ratios (smaller than those analysed here) we expect an isotropic-solid
demixing region in the phase diagram. Columnar as well as solid phases
become relevant in systems of platelets with non-vanishing thickness
at sufficiently high densities. Hence, it is possible that
nematic-nematic demixing in mixtures with high size ratio will be
metastable with respect to columnar ordering. Recent experiments by
Lekkerkerker {\it et al.} on mixtures of charged gibbsite platelets
and silica spheres \cite{doi:10.1021/la101891e,C1SM06535A} have shown
a large isotropic-columnar coexisting region. Columnar ordering has
also been studied theoretically in mixtures of board-like platelets
and spheres in a cubic lattice \cite{C0JM01692F}. Smectic phases are
not very common in discotic liquid crystals, but they could appear in
real systems with finite thickness \cite{PhysRevE.80.041704}. Recent
experiments with mixtures of colloidal gibbsite platelets and spheres
\cite{C0SM01206H} show that also glass formation can occur at
sufficiently high sphere concentrations.

As our theory is mean-field in character, we expect the location of
nematic-nematic critical point to change upon including neglected
fluctuations, say in computer simulations. Such (future) work could
ascertain the stability of nematic-nematic demixing. Furthermore it
would be interesting to consider the stability of the nematic-nematic
transition upon altering the platelet-polymer interactions, in order
to take into account penetrability of platelets and polymers.

The results presented in this work can form the basis for studies of
inhomogeneous platelet-sphere mixtures using DFT, such as the analysis
of the effects of gravity \cite{Wensink2004}, as well as wetting and
confinement effects.


\end{document}